\documentclass[conference]{IEEEtran}

\ifCLASSINFOpdf
\usepackage{url}
\usepackage{cite}
\usepackage{csquotes}
\usepackage{graphicx} 
\usepackage{subcaption}
\usepackage{hhline} 
\usepackage{array}
\usepackage{multirow}
\usepackage{mathtools}
\usepackage{booktabs}
\usepackage{float}
\usepackage{orcidlink}
\usepackage{algorithm2e}
\usepackage{tikz}
\usetikzlibrary{arrows.meta, positioning}
\usepackage{amsfonts,verbatim}

\usepackage{eso-pic}

\else

\fi

\newcommand{\IEEEcopyrightnotice}{
\color{gray}
\textcopyright~2026 IEEE. Personal use of this material is permitted. Permission from IEEE must be obtained for all other uses, in any current or future media, including reprinting/republishing this material for advertising or promotional purposes, creating new collective works, for resale or redistribution to servers or lists, or reuse of any copyrighted component of this work in other works.
\smallskip\par
This paper has been accepted for publication in the Proceedings of the 2026 IEEE Radar Conference (RadarConf26).
}

\begin{document}

\title{
Transformer-Based Heartbeat Monitoring with FMCW Radar Under Random Body Motion
}

\author{
\IEEEauthorblockN{
Matteo Pardi\,\orcidlink{0009-0007-0615-6291}\IEEEauthorrefmark{2}\IEEEauthorrefmark{1},
Amir Hosein Oveis\,\orcidlink{0000-0002-6057-8459}\IEEEauthorrefmark{1},
Saba Kharabadze\,\orcidlink{0000-0003-4504-4181}\IEEEauthorrefmark{1},
Ajeet Kumar\,\orcidlink{0000-0002-7758-8055}\IEEEauthorrefmark{1},
}
\vspace{3pt}
\IEEEauthorblockA{\IEEEauthorrefmark{2}Dept. of Information Engineering, University of Pisa, Pisa, Italy}
\IEEEauthorblockA{\IEEEauthorrefmark{1}Radar and Surveillance Systems (RaSS) National Laboratory, CNIT, Pisa, Italy}
\\
\href{mailto:matteo.pardi@phd.unipi.it}{\texttt{matteo.pardi@phd.unipi.it}} \\
{\texttt{\{a.oveis, saba.kharabadze, ajeet.kumar\}@cnit.it}}
}



\AddToShipoutPictureFG*{
  \AtPageLowerLeft{
    \put(\LenToUnit{(\paperwidth-\textwidth)/2 - 0.05in},\LenToUnit{0.30in}){
      \parbox[b]{\textwidth}{
        \footnotesize\IEEEcopyrightnotice
      }
    }
  }
}

\maketitle

\begin{abstract}
Millimeter-wave Frequency Modulated Continuous Wave (FMCW) radar enables contactless cardiac monitoring, but heartbeat estimation becomes challenging when respiration and random body motion (RBM) distort the radar signal. In this paper, we propose a hybrid framework for 77 GHz FMCW radar that combines model-based signal processing with a Convolutional Neural Network (CNN)–Transformer network. The first block extracts chest displacement and constructs meaningful high-level motion features from raw radar data, while the second block reconstructs a photoplethysmography (PPG)-like signal from the extracted features. In this study, a synchronized PPG signal is used as the ground truth for heartbeat monitoring in supervised training. The method is evaluated following the IEEE AESS Radar Challenge Problem I protocol using the official datasets and figures of merit across three motion scenarios: stationary, deep breathing, and RBM. Results show that the proposed architecture reliably reconstructs the PPG signal in all scenarios, achieving high fidelity in controlled conditions and maintaining robust performance under motion. This enables reliable average heart rate (AHR) and heart rate variability (HRV) estimation even where benchmark methods fail, and leads to the highest total score among the compared approaches.
\end{abstract}

\begin{IEEEkeywords}
mmWave FMCW radar, Non-contact heartbeat monitoring, Random body motion (RBM), CNN--Transformer, Heart rate variability (HRV), Range Migration, PPG.
\end{IEEEkeywords}

\IEEEpeerreviewmaketitle

\section{Introduction}\label{Sec: Introduction}

Non-contact vital sign monitoring is receiving increasing attention because healthcare systems require safe and comfortable long-term solutions. Contact-based devices, such as ECG electrodes and wearable sensors, must be attached to the body and can cause discomfort or limit mobility, especially during long monitoring periods. These limitations are critical for vulnerable groups, such as newborns, elderly patients, and intensive care patients. For this reason, contactless monitoring is attractive for both clinical and home-care applications.

Radar sensing is a promising solution because it detects small chest movements caused by respiration and heartbeat by analyzing the phase of the reflected signal. It offers high spatial resolution, micrometer-level sensitivity, and works through clothes and bedding, independent of lighting conditions. Moreover, the recent approval \cite{Approval_in_europe_ETSI2008, Commission_Rule_76_81_allowed} of the 76–81 GHz band and advances in radar-on-chip technology have enabled compact and low-cost systems. In this work, we focus on heartbeat monitoring using a single mmWave FMCW radar.

\begin{figure}[t]
\centering
\includegraphics[width=\columnwidth]{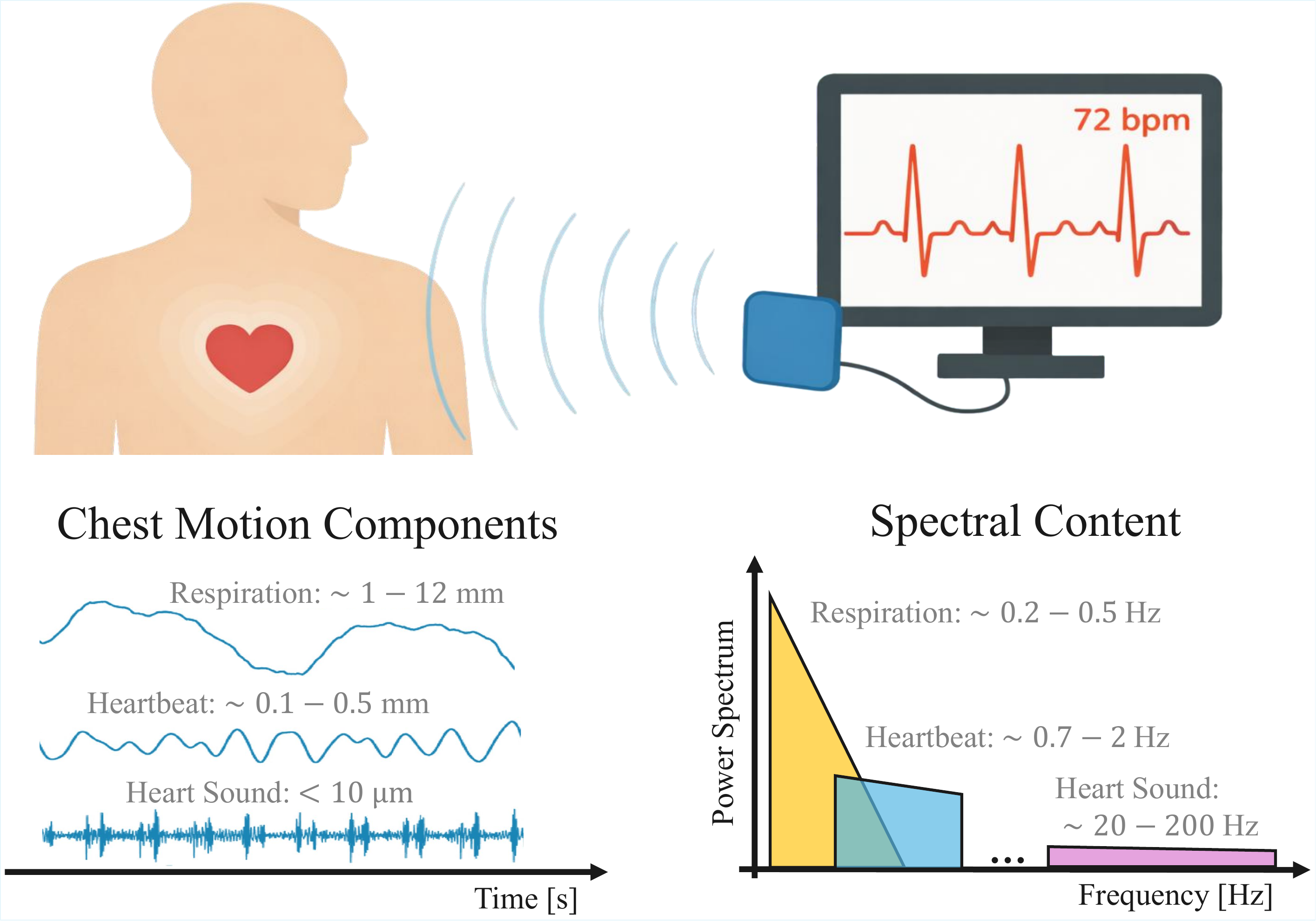}
\caption{Overview of the chest-motion information embedded in the radar signal in the time and frequency domains. Values are taken from~\cite{rong_noncontact_2024, Rong_acoustic_2026_IEEE_IOT}.}
\label{fig:heartbeat-monitoring}
\end{figure}

However, achieving reliable radar-based heartbeat monitoring remains challenging. The cardiac signal is very weak and must be separated from stronger motion components (see Figure \ref{fig:heartbeat-monitoring}), especially respiration and random body movement (RBM). While respiration suppression has been widely studied, RBM remains a major open issue because it can easily mask heartbeat-related displacements.

Several methods have been proposed for single-sensor radar-based heartbeat monitoring \cite{zhang_overview_2023}. These include heart sound analysis \cite{rong_noncontact_2024, Rong_acoustic_2026_IEEE_IOT}, spectral techniques to reduce respiratory harmonics \cite{tu_fast_2016, ahmad_vital_2018}, adaptive wavelet methods \cite{dai_enhancement_2022}, probabilistic approaches such as hidden Markov models \cite{xia_radar-based_2021}, model-driven decomposition methods like VMD, A-VMD and EMD\cite{Hao_Nature_FMCW_A_VMD,Zhang_Nature_Matrix_Coeff}, and machine learning strategies, from linear models \cite{saluja_supervised_2020} to CNNs \cite{yuan_heart_2024}, LSTMs \cite{shi_segmentation_2019}, and Transformers \cite{chen_multi-feature_2025}. Although these approaches show promising results, robustness to RBM in single-sensor systems is still limited, and many methods are not suitable for real-time practical use.

In this paper, we propose a hybrid supervised learning approach that combines model-based signal processing and deep learning. First, raw radar data are processed to estimate chest displacement. From this signal, we extract physically meaningful handcrafted features related to respiration, heartbeat, heart sound, and coarse RBM indicators. These features are then used as input to a CNN--Transformer network. The CNN captures local temporal patterns, while the Transformer models long-range dependencies and focuses on more reliable signal segments. The network is trained to reconstruct the PPG signal, enabling heartbeat estimation. Supervised learning is adopted because cardiac morphology changes across subjects and time, making analytical modeling difficult, while labeled data can be easily collected using ECG or PPG as reference. By combining handcrafted features with deep learning, the proposed method improves interpretability and robustness.

The main contributions of this paper are as follows:
\begin{itemize}
    \item A signal processing pipeline to estimate chest displacement from raw radar data, accounting for range migration.
    \item Definition of a set of physically meaningful handcrafted features derived from the displacement signal, suited for PPG-like reconstruction via deep learning.
    \item A CNN--Transformer architecture for estimating the PPG signal from the extracted features.
\end{itemize}

This paper is organized as follows. Section~II presents the proposed methodology. Section~III describes the dataset and the experimental setup. Section~IV reports and discusses the results. Finally, Section~V concludes the paper.

\section{Proposed Methodology}

\begin{figure*}[t]
\centering
\includegraphics[width=\textwidth]{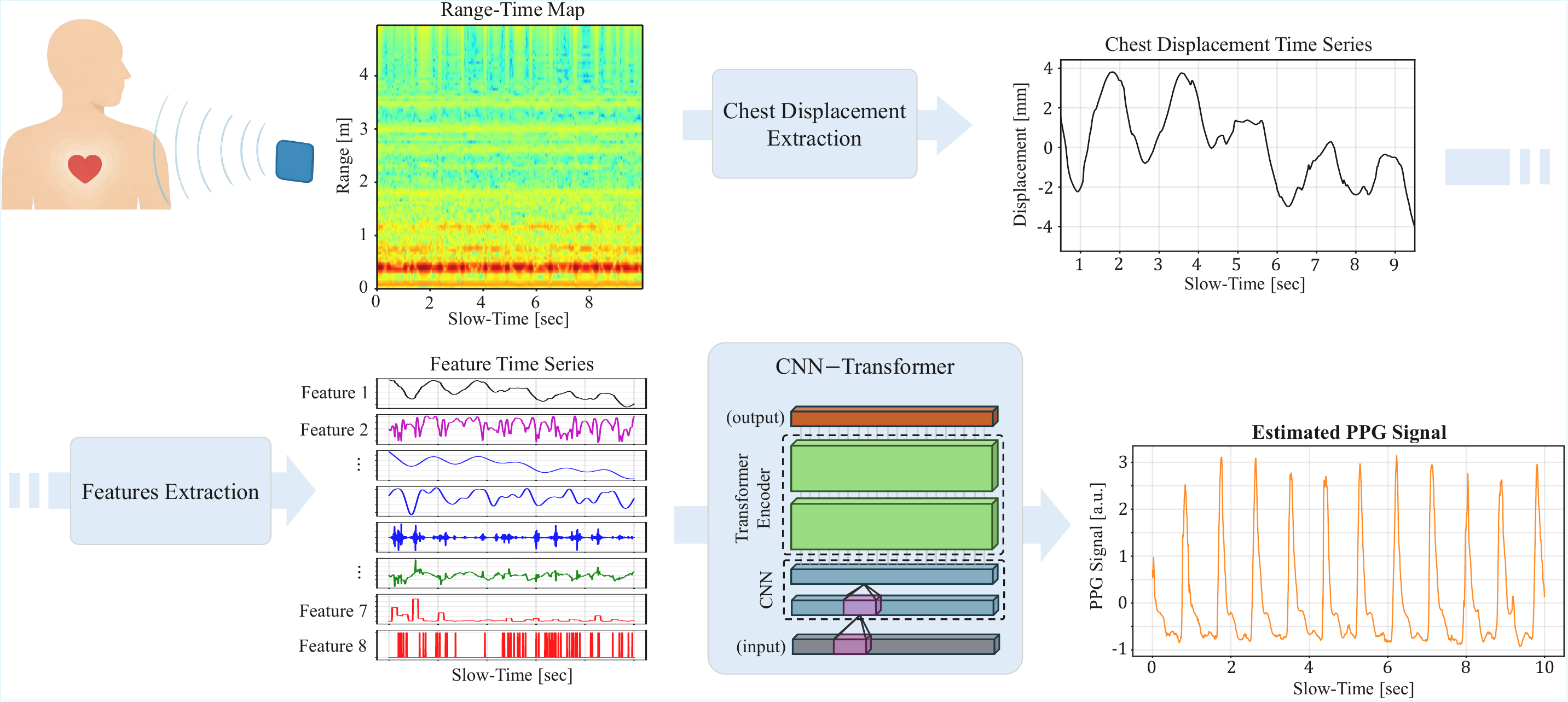}
\caption{End-to-end scheme of the proposed method. Starting from the radar range--time matrix, the chest range bin is tracked and the signal phase is extracted, yielding a chest-displacement time series. A selected set of handcrafted features is then derived and processed by a CNN--Transformer network to estimate the PPG signal, from which heartbeat monitoring is straightforward.}
\label{fig:algorithm-overview}
\vspace{-0.5mm}
\end{figure*}

The proposed method consists of two main components. The first is a model-based signal processing block that converts raw mmWave FMCW radar measurements into physiologically meaningful features (time series). The second is a CNN--Transformer network that processes the extracted features to reconstruct the PPG signal. A high-level end-to-end overview of the method is shown in Figure \ref{fig:algorithm-overview}, and described in detail next.

\subsection{Radar Signal Processing Block}
\label{subsec:radar_signal_processing}

\begin{figure}[t]
\centering
\includegraphics[width=0.8\columnwidth]{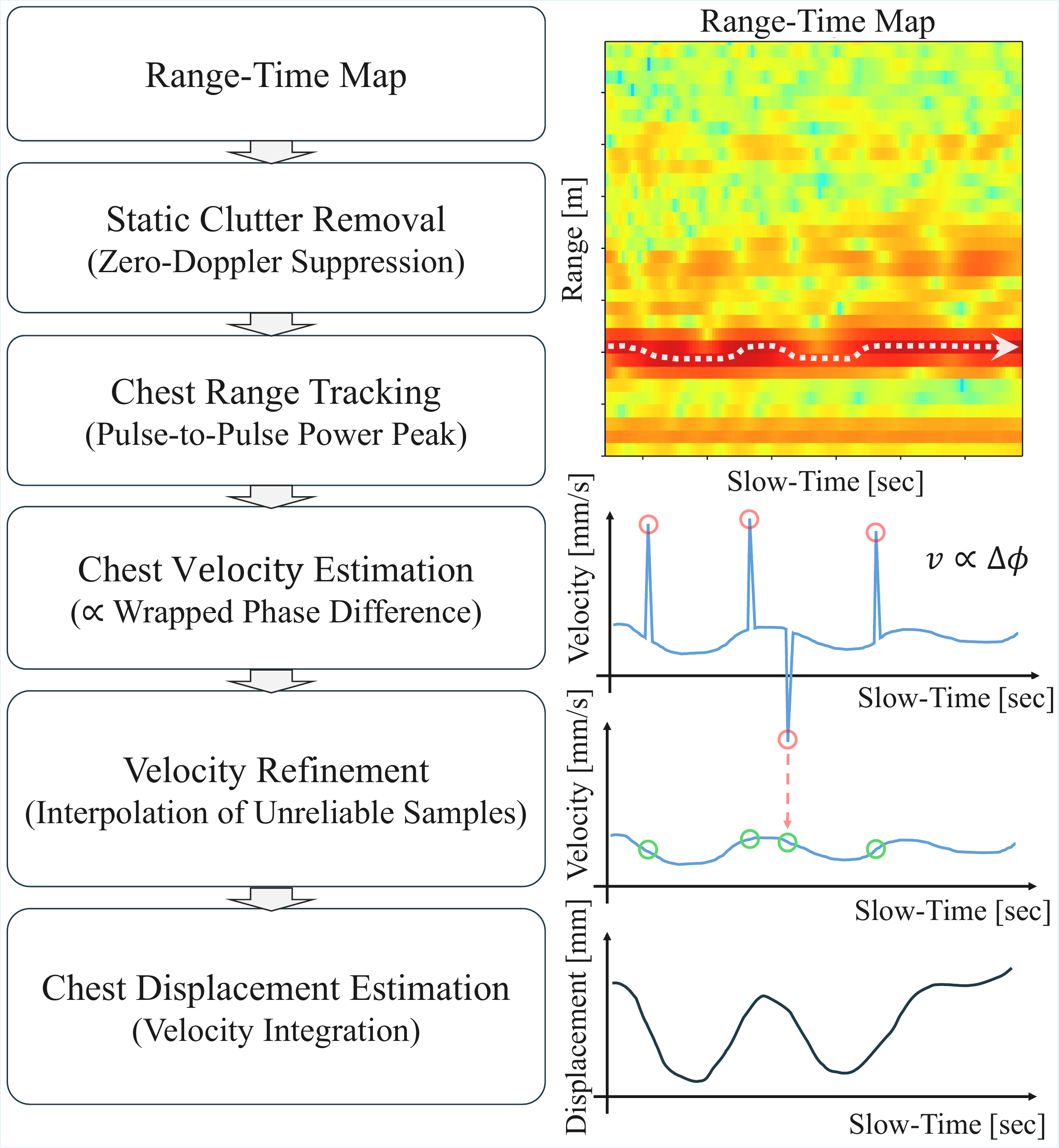}
\caption{Schematic of the signal processing steps used to extract the chest-displacement time series from the raw radar data.}
\label{fig:from-radar-to-displacement}
\vspace{-1.5mm}
\end{figure}

A schematic of the signal processing block is illustrated in Figure \ref{fig:from-radar-to-displacement}. Let the complex radar data acquired over a CPI be denoted as
\begin{equation}
\mathbf{X} \in \mathbb{C}^{K \times T},
\end{equation}
where $K$ is the number of fast-time frequency (range) samples and $T$ is the number of slow-time frames within the CPI of duration $T_{\mathrm{CPI}}$. In our implementation, $T_{\mathrm{CPI}} = 10\,\mathrm{s}$.

\subsubsection{Range Processing and Clutter Removal}
Each CPI is first multiplied by a Hamming window (to suppress sidelobes) along the fast-time dimension and a one-dimensional FFT is then applied to transform the data from the frequency--time domain to range--time profiles. Static clutter and stationary reflections are suppressed by subtracting the mean value across the slow-time dimension for each range bin, effectively removing the zero-Doppler component.

\subsubsection{Chest Range Bin Selection}
For each slow-time frame, the dominant chest reflection is identified by selecting the range bin corresponding to the maximum signal power. This adaptive selection enables the tracking of range variations induced by subject motion. 
After selecting the dominant chest range bin, the extracted slow-time complex signal can be modeled as
\begin{equation}
S[t] = A[t] e^{j\phi[t]} + n[t],
\end{equation}
where $t$ denotes the slow-time index, $A[t]$ is the amplitude, and $n[t]$ represents additive noise. 
The phase $\phi[t]$ is directly related to chest displacement through
\begin{equation}
\phi[t] = \frac{4\pi}{\lambda} R[t],
\end{equation}
where $\lambda$ is the radar wavelength and $R[t]$ denotes the composite chest displacement containing respiration, heartbeat, heart sound and random body motion components (see Figure \ref{fig:heartbeat-monitoring}).

\subsubsection{Phase, Amplitude, and Velocity Estimation}
From the extracted chest IQ signal, the instantaneous amplitude $A[t]$ and wrapped phase are computed. Phase differences between consecutive frames $\Delta \phi$ are calculated and wrapped to the interval $[-\pi, \pi)$. These phase differences are converted into instantaneous radial velocity as $v[t] = \frac{\lambda \Delta \phi [t]}{4 \pi} \mathrm{PRF}$, where $\mathrm{PRF}$ is the frame rate.

\subsubsection{Velocity Refinement}
Velocity samples affected by range bin migration are identified as non-reliable and are replaced using linear interpolation over neighboring reliable samples.

\subsubsection{Displacement Reconstruction}
The corrected velocity $\hat{v}[t]$ is integrated over slow time to obtain chest displacement $\hat{R}[t]$. The resulting displacement signal is centered around zero by removing its mean value.

\subsubsection{Feature Construction}
\begin{figure}[t]
\centering
\includegraphics[width=0.9\columnwidth]{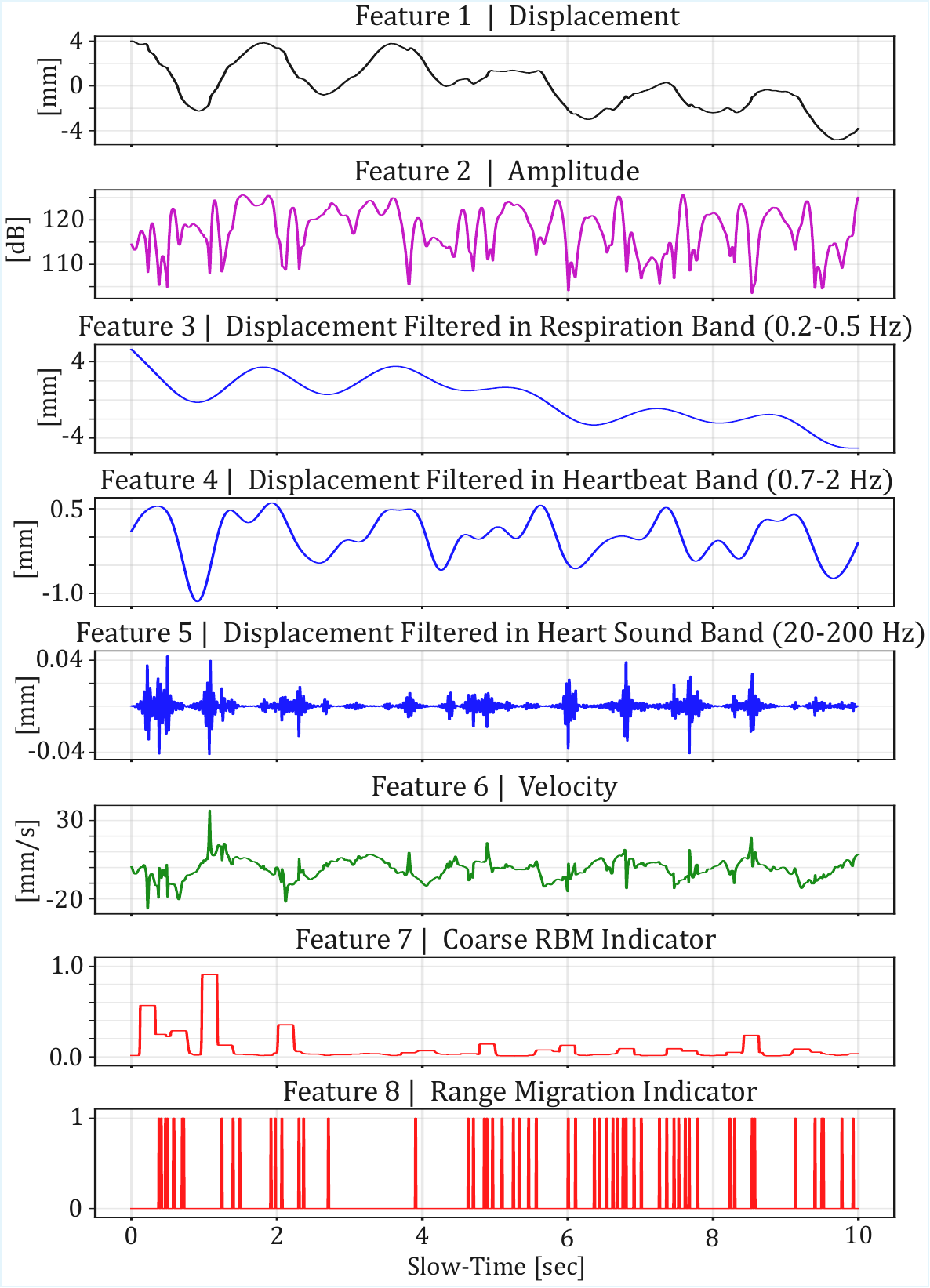}
\caption{Feature representation obtained from a sample CPI. This eight-dimensional time series serves as the input to the CNN--Transformer network for PPG signal estimation.}
\label{fig:features}
\end{figure}

From the processed slow-time signals, a set of eight complementary feature channels is constructed for each CPI (see Figure \ref{fig:features}):
\begin{itemize}
    \item Chest displacement $\hat{R}[t]$;
    \item Chest signal amplitude $A[t]$ in dB;
    \item Band-pass filtered $\hat{R}[t]$ in respiration band (0.2--0.5~Hz);
    \item Band-pass filtered $\hat{R}[t]$ in heartbeat band (0.7--2.0~Hz);
    \item Band-pass filtered $\hat{R}[t]$ in heart sound band (20--200~Hz);
    \item Instantaneous chest velocity $\hat{v}[t]$;
    \item Coarse RBM activity indicator: soft $(0, 1)$ signal from $\mathrm{Sigmoid}(|\hat{v}[t]| - v_\mathrm{th}, \sigma_v)$, where $\mathrm{Sigmoid}(x, a) = 1/(1 + \exp(-x/a))$, with $v_\mathrm{th}=25$~mm/s and $\sigma_v=5$~mm/s, followed by a max-filter with 0.2~s window;
    \item Range migration indicator: binary flag set to 1 at pulses where the dominant range bin index changes with respect to the previous or next pulse.
\end{itemize}
All band-pass filters are implemented as linear-phase FIR filters and applied using forward--backward filtering and odd-symmetric padding at the signal boundaries. The heart sound filter is implemented as a $>20$ Hz high-pass filter, given the 200 Hz slow-time sampling rate (Nyquist is 100 Hz).

\subsubsection{Output Representation}
The resulting feature tensor for each CPI is given by
\begin{equation}
\mathbf{F} \in \mathbb{R}^{C \times T},
\end{equation}
where $C = 8$ denotes the number of extracted feature channels. These feature sequences preserve the full slow-time resolution and are directly provided as input to the subsequent Transformer-based model.

\subsection{Transformer-Based Heartbeat Monitoring}
\label{subsec:transformer_heartbeat}

The feature tensor $\mathbf{F} \in \mathbb{R}^{C \times T}$ is fed into a Transformer-based network to estimate the PPG signal $\hat{\mathbf{y}} \in \mathbb{R}^{T}$. In other words, the network maps a $C$-channel input sequence to a single-channel output sequence. Input features and target PPG are z-scored using the training set mean and standard deviation, and predictions are de-normalized at inference. The model consists of four blocks: a convolutional stem, a positional-encoding layer, a Transformer encoder, and a convolutional output head.

\paragraph{Convolutional Stem}
The input tensor $\mathbf{F}$ is first processed by a one-dimensional ReLU convolutional stem that maps the feature space to a latent embedding of dimension $d_{\text{model}}$. The stem preserves temporal resolution and produces an intermediate representation
\begin{equation}
\mathbf{Z} \in \mathbb{R}^{d_{\text{model}} \times T},
\end{equation}
which performs local temporal feature extraction and channel mixing across the multistream inputs.

\paragraph{Positional Encoding}
The embedded features $\mathbf{Z}$ are augmented with standard additive sinusoidal positional encoding, yielding
\begin{equation}
\mathbf{Z}' \in \mathbb{R}^{d_{\text{model} \times T}}.
\end{equation}
The positional encoding allows the model to retain temporal ordering information.

\paragraph{Transformer Encoder}
Temporal dependencies are modeled using a Transformer encoder composed of $N_\mathrm{layers}$ encoder layers. Each layer employs multi-head self-attention with $N_\mathrm{heads}$ attention heads and a position-wise feed-forward network. Residual connections, layer normalization, and dropout are applied following standard Transformer design principles. ReLU is used. The encoder preserves the embedding dimensionality, producing an output
\begin{equation}
\mathbf{H} \in \mathbb{R}^{d_{\text{model}} \times T}.
\end{equation}

\paragraph{Output Head}
The Transformer output is mapped to a scalar waveform using a single 1×1 Conv1D, producing the estimated PPG signal at each slow-time step:
\begin{equation}
\hat{\mathbf{y}} \in \mathbb{R}^{T}.
\end{equation}

The main architectural hyperparameters, such as $d_{\text{model}}$, $N_\mathrm{layers}$, $N_{heads}$, together with the design of the convolutional stem, are selected through systematic validation and reported in Section~IV.

\section{Data Description and Experimental Setup}

The experimental dataset was provided by the IEEE AESS Challenge Problem I \cite{Rong2025RadarChallenge}. Data were collected using a 77 GHz mmWave FMCW radar system based on the Texas Instruments platform (AWR2243BOOST or AWR1843BOOST) with a DCA1000 capture card.

Table \ref{table:hardware-description} summarizes the main radar specifications. The system employs three transmit and four receive antennas, forming a 12-element virtual array via TDM-MIMO. A bandwidth of 3.6 GHz provides a range resolution of approximately 4.2 cm. Each transmit antenna radiates about 13 dBm, enabling low-power operation suitable for close-range physiological monitoring. The ADC sampling rate is 2 MSps, and the frame rate is 200 Hz.

\begin{table}[t]
    \centering
    \caption{Radar system specifications.}
    \label{table:hardware-description}
    \begin{tabular}{|l|l|}
    \hline
    \textbf{Parameter} & \textbf{Specification} \\ \hline
    Operating Frequency & 77.0--80.6 GHz \\ \hline
    Radar Type & FMCW mmWave Radar \\ \hline
    Number of Transmit Antennas & 3 \\ \hline
    Number of Receive Antennas & 4 \\ \hline
    Virtual Antennas (TDM-MIMO) & 12 (3 Tx × 4 Rx) \\ \hline
    Bandwidth & 3.6 GHz \\ \hline
    Chirp Duration & 60 $\mu$s \\ \hline
    Transmit Power & 13 dBm (per Tx) \\ \hline
    Range Resolution & 4.2 cm (with 3.6 GHz BW) \\ \hline
    ADC Sampling Rate & 2 MSps \\ \hline
    Frame Rate & 200 Hz \\ \hline
    \end{tabular}
\end{table}

Approximately 45 minutes of data were collected under three conditions: stationary, deep breathing, and random body movement. Each condition includes three independent recordings of about 5 minutes, yielding nine recordings in total. These scenarios evaluate heartbeat monitoring robustness under progressively increasing motion interference, particularly RBMs encountered in practical environments.

During acquisition, one chirp was transmitted per frame. The recorded data consist of complex ADC samples obtained after dechirping. The receive channels were summed to focus on the chest, producing a two-dimensional data cube of fast-time samples and slow-time frames. Each file contains 100 fast-time samples per chirp and 60,000 slow-time frames.

Ground-truth cardiac activity was provided by a synchronized PPG signal acquired using a pulse oximeter and stored within each data file. The PPG was resampled to 200 Hz to match the radar slow-time rate. For each recording, we normalized the PPG signal to have zero mean and unit standard deviation, ensuring all data are on a consistent standardized scale suitable for supervised learning.

For training and evaluation, each recording was split into 70\%, 15\%, and 15\% subsets for training, validation, and testing, respectively. This within-recording split ensures representation of all three scenarios in each phase. Performance metrics were computed on the aggregated test portions. Although a scenario-level or recording-level split would better prevent potential data leakage, the limited number of recordings required this strategy to maintain sufficient variability across sets.

\section{Results and Discussion}

\subsection{Performance Metrics}
Performance is evaluated according to the IEEE AESS Challenge Problem~I \cite{Rong2025RadarChallenge} protocol, which jointly assesses Average Heart Rate (AHR) and Heart Rate Variability (HRV).

\subsubsection{Average Heart Rate (AHR)}
The estimated and reference PPG signals are processed using a short-time Fourier transform (STFT) with a 10-s Hann window and 1-s step, corresponding to the CPI duration and overlap defined in the evaluation protocol. For each STFT frame, the dominant spectral peak of the magnitude spectrum is selected, and its frequency is converted to beats per minute (BPM), yielding one AHR estimate per CPI. 
The AHR error ($\hat{\delta}_{\mathrm{AHR}}$) is quantified using the root-mean-square error (RMSE)
between the estimated and reference heart rates. The corresponding figure of merit (FoM) is defined
as the AHR RMSE normalized by the CPI duration:
\begin{equation}
\mathrm{FoM}_{\mathrm{AHR}} = \frac{1}{T_{\mathrm{CPI}} \cdot \hat{\delta}_{\mathrm{AHR}}}.
\end{equation}

\subsubsection{Heart Rate Variability (HRV)}
HRV performance is evaluated in the time domain using beat-to-beat cardiac intervals (commonly referred to as RR intervals) extracted from the estimated and reference PPG signals. Cardiac peaks in the estimated signal are detected using a peak-finding algorithm with a minimum inter-peak distance of 0.5 s (corresponding to a maximum heart rate of 120 BPM) and an adaptive minimum prominence threshold equal to the PPG signal standard deviation. Successive peak locations are used to compute RR intervals, which are then converted to instantaneous heart rate values. 
The primary error metric is the RMSE between the estimated and reference heart rate sequences ($\hat{\delta}_{\mathrm{HRV}}$). In addition, the Pearson correlation coefficient between the estimated and reference signals ($C_h$) is computed to account for waveform similarity and potential beat detection mismatches. The HRV figure of merit is defined as
\begin{equation}
\mathrm{FoM}_{\mathrm{HRV}} = \frac{C_h}{\hat{\delta}_{\mathrm{HRV}}}.
\end{equation}

\subsubsection{Overall Score}
The final score is obtained as a weighted combination of the AHR and HRV figures of merit
and averaged across all datasets:
\begin{equation}
\label{eq:score}
\mathrm{SCORE} = (100 \times 10)\cdot \mathrm{FoM}_{\mathrm{AHR}} + \left(\frac{100}{60 \times 0.9}\right)\cdot \mathrm{FoM}_{\mathrm{HRV}}.
\end{equation}

For $D$ datasets, the total score is given by
\begin{equation}
\mathrm{SCORE}_{\mathrm{Total}} = \frac{1}{D}\sum_{i=1}^{D} \mathrm{SCORE}[\text{dataset } i].
\end{equation}

\subsection{Hyperparameter Optimization}
\textcolor{black}{To identify a well-performing configuration, we performed a systematic hyperparameter search over the space summarized in Table~\ref{tab:hyperparam-space} using Ray Tune \cite{RayTune}. In total, we trained roughly 300 models, organized into seven experimental campaigns. Each campaign comprised multiple Ray Tune trials (i.e., individual model training jobs), with up to four trials trained in parallel; across all 300 trials, the median runtime was about 9 minutes, largely because the Asynchronous Successive Halving Algorithm \cite{asha} early-stopping criterion terminated underperforming configurations early, whereas for the subset of trials that ran to completion the median runtime was about 65 minutes. Initial random sampling explored the space, followed by refinement with the Tree-structured Parzen Estimator (TPE) \cite{hyperopt}, which handles mixed discrete and continuous search spaces efficiently.}

\begin{table}[t]
\centering
\caption{Hyperparameter search space.}
\label{tab:hyperparam-space}
\begin{tabular}{lll}
\toprule
\textbf{Parameter} & \textbf{Type} & \textbf{Range / Values} \\
\midrule
\multicolumn{3}{l}{\textit{Training}} \\
Learning rate   & Log-uniform & $[10^{-5},\; 10^{-2}]$ \\
Batch size      & Categorical & \{4, 8, 16, 64\} \\
Weight decay    & Log-uniform & $[10^{-5},\; 10^{-3}]$ \\
Optimizer       & Categorical & \{Adam, AdamW, SGD\} \\
\midrule
\multicolumn{3}{l}{\textit{Model architecture}} \\
$d_{\text{model}}$  & Categorical & \{64, 96, 128\} \\
Attention heads     & Categorical & \{2, 4, 8\} \\
Encoder layers      & Categorical & \{1, 2, 3, 4\} \\
Dropout             & Uniform     & $[0.05,\; 0.2]$ \\
CNN stem config     & Categorical & 6 architectures$^{\ast}$ \\
\bottomrule
\multicolumn{3}{l}{\footnotesize $^{\ast}$Varying number of layers, channel widths, and kernel sizes.}
\end{tabular}
\vspace{-0.5mm}
\end{table}

\textcolor{black} {We also evaluated simple data-augmentation schemes, including Markov-based random-walk strategies for simulating synthetic RBM, with no validation improvement. The best-performing configuration uses AdamW (betas=(0.9, 0.999), eps=1e-08) with learning rate $9.2\times10^{-4}$, batch size 4, weight decay $1.3\times10^{-5}$, MSE loss, embedding dimension $d_{\text{model}}=96$, 8 attention heads, 2 encoder layers, dropout 0.16, and a three-layer CNN stem with channel progression $[32,64,d_{\text{model}}]$ and kernel size 3. This model was trained for the duration of 40 epochs.}

\subsection{Final model and discussion}

\begin{figure}[t]
    \centering
    \includegraphics[width=0.95\columnwidth]{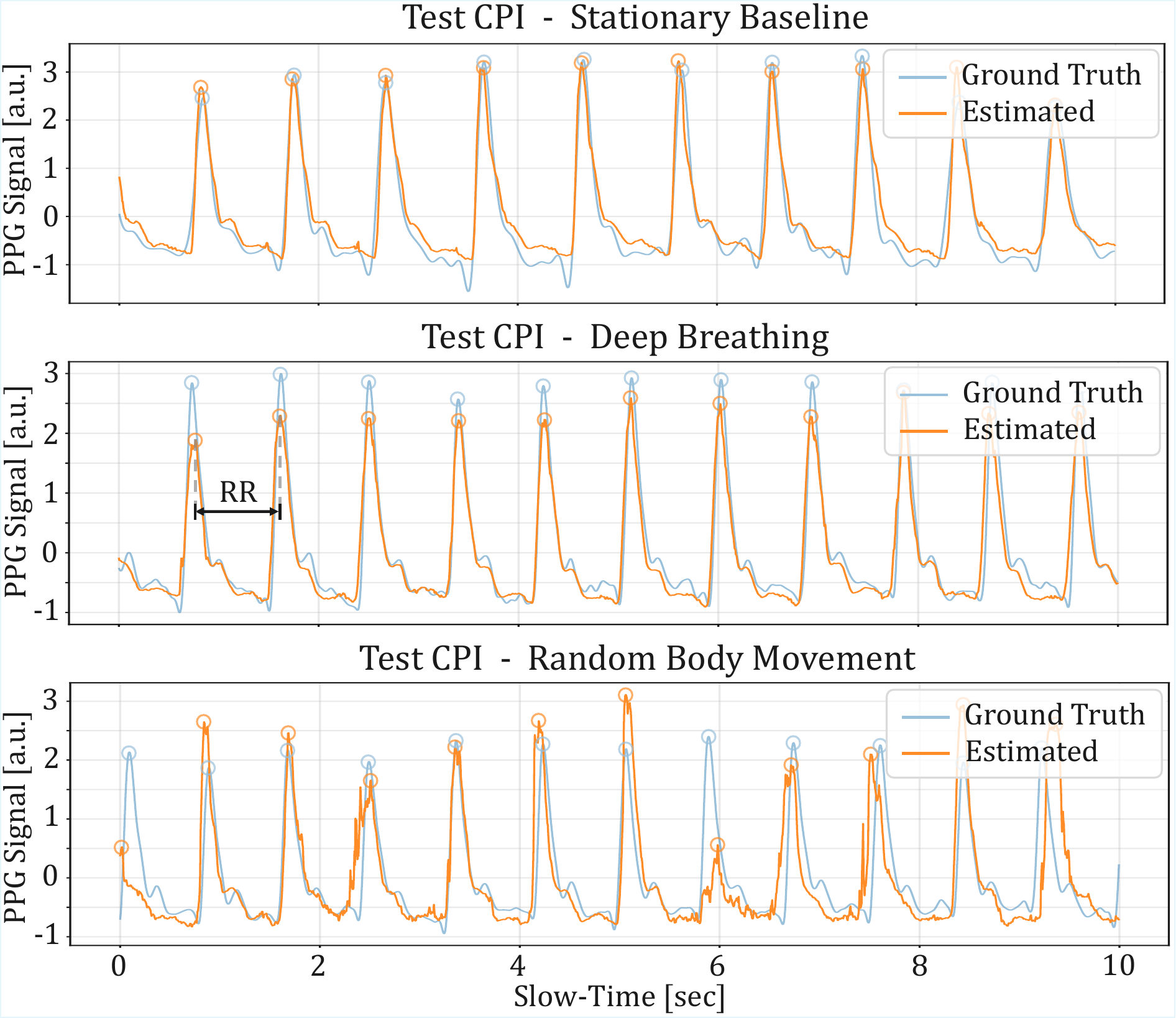}
    \caption{Estimated PPG signal (CNN--Transformer output) compared with the ground truth. Three representative CPIs from the test set are shown, corresponding (from top to bottom) to the stationary baseline, deep breathing, and RBM datasets.}
    \label{fig:output-example}
    \vspace{-1.5mm}
\end{figure}

Table~\ref{tab:score-comparison} summarizes the results and compares our approach with the benchmark methods from the IEEE AESS Challenge Problem~I \cite{Rong2025RadarChallenge}: Filtering, Blind Source Separation, and HF-Heartbeat. Figure~\ref{fig:output-example} shows some representative CPIs of the PPG signal estimated by the proposed method.

\begin{table*}[t]
    \centering
    \caption{Performance comparison against the methods in~\cite{Rong2025RadarChallenge} using the official challenge figures of merit and scores. The values of $\mathrm{FoM}_{\mathrm{AHR}}$ and $\mathrm{FoM}_{\mathrm{HRV}}$ already include the multiplicative factor defined in~(\ref{eq:score}), as in~\cite{Rong2025RadarChallenge}.}
    \label{tab:score-comparison}
    \setlength{\tabcolsep}{6pt}
    \renewcommand{\arraystretch}{1.25}
    \begin{tabular}{lcc|c|cc|c|cc|c|c}
    \toprule
    & \multicolumn{3}{c|}{\textbf{Stationary Baseline}}
    & \multicolumn{3}{c|}{\textbf{Deep Breathing}}
    & \multicolumn{3}{c|}{\textbf{Random Body Movement}}
    & \textbf{Total Score} \\
    \textbf{Method}
    & FoM$_{\mathrm{AHR}}$ & FoM$_{\mathrm{HRV}}$ & Score
    & FoM$_{\mathrm{AHR}}$ & FoM$_{\mathrm{HRV}}$ & Score
    & FoM$_{\mathrm{AHR}}$ & FoM$_{\mathrm{HRV}}$ & Score
    &  \\
    \midrule
    Filtering
    & 13.2 & 4.0 & 17.2
    & 3.5 & 0.8 & 4.3
    & 2.6 & 0.0 & 2.6
    & 8.0 \\
    Blind Source Separation
    & 13.2 & 1.8 & 15.0
    & 3.7 & 1.1 & 4.8
    & 3.0 & 0.0 & 3.0
    & 7.6 \\
    HF-Heartbeat
    & 255.4 & \textbf{106.6} & \textbf{362.1}
    & 115.7 & 12.5  & 128.1
    & 3.4   & 0.4   & 3.8
    & 164.7 \\
    Our Method
    & \textbf{266.2} & 52.0 & 318.2
    & \textbf{227.6} & \textbf{65.7} & \textbf{293.2}
    & \textbf{171.7} & \textbf{23.9} & \textbf{195.6}
    & \textbf{269.0} \\
    \bottomrule
    \end{tabular}
\end{table*}

In the ideal stationary case, the proposed method achieves a good performance, even if it does not reach the best performance and stays below the state-of-the-art method HF-Heartbeat, especially for the HRV metric. However, when motion is present, it clearly outperforms all the others. This holds for deep breathing and even more for random body movement, where the other methods perform poorly while ours still achieves good performance and degrades gracefully.

The estimated PPG in Figure~\ref{fig:output-example} shows a high-quality reconstruction of the true PPG for the stationary baseline and deep breathing cases. For random body movement, the estimate still follows very well the overall trend of the true PPG, though with some loss in reconstruction quality.

Overall, these results show that the proposed framework is more robust than the other methods when conditions are non-ideal or affected by unwanted motion.

\section{Conclusion}

The results show that the proposed architecture reliably reconstructs the PPG signal in all tested scenarios, achieving high fidelity in controlled conditions and maintaining robust performance under motion-corrupted settings. In stationary and deep breathing conditions, the signal morphology is well preserved, including peak shape and timing. Even with RBMs, the model follows the overall signal trend and correctly detects the main peaks. Performance degrades gradually under motion, indicating robustness in non-ideal conditions.

For metrics such as AHR and HRV, the proposed method clearly outperforms the IEEE AESS Challenge Problem I benchmark techniques. In the RBM scenario, it provides reliable estimates of both metrics, while the benchmark methods (Filtering, Blind Source Separation, and HF-Heartbeat) fail completely under motion-corrupted conditions. This confirms the practical value of the proposed architecture for radar-based vital sign monitoring in realistic environments.

Future work will focus on:
\begin{itemize}
    \item Testing other neural networks to improve the balance between performance and computational cost, with attention to lightweight and real-time implementations.
    \item Validating the method on a larger and more diverse dataset.
\end{itemize}

\section{Acknowledgment}
This work was conducted as part of the IEEE Aerospace and Electronic Systems Society (AESS) Challenge Problem I, entitled \textit{Radar-Based Heartbeat Monitoring in Dynamic Scenarios}. The challenge is financially supported by the IEEE AESS. The research activities were carried out at the RaSS National Laboratory, CNIT, Pisa, Italy.

\bibliographystyle{IEEEbib}
\bibliography{ref}
\end{document}